\documentclass[5p,final,authoryear]{elsarticle}

\usepackage[T1]{fontenc}
\usepackage[utf8]{inputenc}
\usepackage[english]{babel}
\usepackage{amsmath,amssymb}
\usepackage{graphicx}
\usepackage{float}
\usepackage{enumitem}
\usepackage[colorlinks=true,allcolors=blue]{hyperref}
\biboptions{round,authoryear}
\begin{document}

\begin{frontmatter}

\title{A Gabor--Epps uncertainty principle for traders}

\author[uct-sta]{Tim Gebbie}
\ead{tim.gebbie@uct.ac.za}
\address[uct-sta]{Department of Statistical Sciences, University of Cape Town, Rondebosch 7701, Western Cape, South Africa}

\begin{abstract}
We propose a Gabor--Epps uncertainty principle for practical trading. The key idea is that high-frequency correlation is not observed in clock time alone, but is resolved through market activity, order-flow overlap, and finite coupling response. This suggests six simple rules of thumb that may be useful to traders and trading programs operating at market-making frequencies, particularly those crossing books and markets below the average human response time. Throughout, the observation window is clock-dependent: in calendar time it is a physical interval, in trade time it is a trade-count interval, and in volume time it is a volume bucket. In summary: at event scales, the more precisely one localises market activity in time, the less well one can resolve stable cross-asset dependence. The more one resolves dependence, the more one has coarse-grained away the event-time structure that generated it. This can generate substantial clock risk.
\end{abstract}


\end{frontmatter}

\tableofcontents

\section{The Gabor analogy}

In Gabor's setting, a signal cannot be localised arbitrarily sharply in both time and frequency \citep{Gabor1946}.  With the usual angular-frequency convention,
\begin{equation}
    \Delta t\,\Delta \omega \geq \frac{1}{2}.
\end{equation}

This is a Fourier and signal-processing statement.  The word uncertainty is used here in that sense, not in the quantum-mechanical sense.  No physical measurement effect or position--momentum analogy is being invoked.  The relevant analogy is with radar, speech, and time--frequency analysis: a finite-window signal cannot be localised arbitrarily sharply in both a clock and its conjugate rate domain.

A short observation window gives good localisation in the chosen clock, but it necessarily blurs the rate content of the signal.  A longer observation window gives better rate resolution, but it averages away more of the event-time structure.  For example, if a market response oscillates or relaxes at about \(2\,{\rm s}^{-1}\), then a \(100\) millisecond window only contains about \(0.2\) response times.  It may locate a queue event, but it has very little chance of resolving that response rate.  A \(0.5\) second window contains about one response time, while a \(2.5\) second window contains about five response times and is much closer to a usable rate measurement.

Strictly, the Gabor bound is stated in terms of second-moment spreads of a windowed signal and its Fourier transform.  In the trading rules below, \(\Delta_C\) is used as an effective window width in the chosen market clock.  The numerical constant \(1/2\) is therefore a convention-dependent resolution constant, not a calibrated financial parameter.  With angular frequency \(\omega\) the bound is written as above; if cycles per second are used instead, the numerical constant changes.  In this paper the financial content enters only after choosing the market signal and the rate one wants to resolve.

A natural market analogue of frequency is an activity or response rate: event intensity, refresh intensity, turnover rate, order-flow reversal frequency, liquidity replenishment rate, cross-asset coupling rate, and so on. These are event or activity rates.  This is a modelling choice, not a new physical law.  It is reasonable when the object being traded is itself a rate-limited market mechanism: shared refresh, order-flow overlap, liquidity replenishment, or cross-book response.  The useful question is then practical and dimensional: how many refresh opportunities or response times lie inside the observation window?  A \(50\) millisecond imbalance statistic may be useful for reacting to the book, but it should not be confused with a mature estimate of a cross-asset coupling rate that unfolds over one or several seconds.

Because several market clocks are possible, it is useful to write the observation window as $\Delta_C$, where $C$ denotes the chosen clock.  Thus $C=t$ denotes calendar time, $C=N$ denotes event or trade time, and $C=V$ denotes volume time.  This follows the alternative-clock view in the Epps literature, where calendar, trade, and volume time need not produce the same correlation-emergence curve \citep{ChangPienaarGebbie2021PhysicaA,AngstmannGebbie2026NonUnique}.  In these cases
\begin{equation}
\begin{aligned}
    \Delta_t &= \text{calendar time interval},\\
    \Delta_N &= \text{event or trade-count interval},\\
    \Delta_V &= \text{volume bucket}.
\end{aligned}
\end{equation}
For a single asset, the event clock may be the asset's own trade count.  For a pair, the event-time clock must be specified by the sampling convention: pooled events, refresh-time sampling, previous-tick synchronisation, or another joint event clock.  Different choices can produce different Epps curves.  The generic Epps-scale window $\Delta$ should therefore be read as $\Delta_C$ once a clock has been chosen.  The dual variable $\omega_C$ is an angular frequency per unit of the chosen clock.  If cycles rather than angular frequency are used, the numerical constant in the Gabor bound changes in the usual way.

The relevant dimensionless products are then of the form
\begin{equation}
    \lambda^{C}_{ij}\Delta_C, \qquad \kappa^{C}_{ij}\Delta_C,
\end{equation}
where $\lambda^{C}_{ij}$ is an effective joint refresh or clock-overlap rate for assets $i$ and $j$ in clock $C$, and $\kappa^{C}_{ij}$ is an effective coupling or relaxation rate in that clock \citep{AngstmannGebbie2026LOB}.  These products count how many market refresh opportunities or coupling-response times are contained in an observation window of size $\Delta_C$.  They can therefore be used to create concrete running estimates of trading opportunities and correlation maturity.  For example, $\kappa^C_{ij}=2\,{\rm s}^{-1}$ and $\Delta_C=0.25\,{\rm s}$ gives $\kappa^C_{ij}\Delta_C=0.5$, which is around the Gabor-resolution threshold.  The same rate with $\Delta_C=2.5\,{\rm s}$ gives $\kappa^C_{ij}\Delta_C=5$, which is closer to an Epps-maturity threshold.

We use the Gabor uncertainty principle directly as a statement about a windowed market signal.  For a chosen clock $C$, let $Y^C_{ij}$ denote the correlation-bearing signal: a return product, signed order-flow co-movement, or order-imbalance response between assets $i$ and $j$.  Observing $Y^C_{ij}$ through a finite window of width $\Delta_C$ imposes a spectral uncertainty $\Delta\omega_C$.  The Gabor inequality gives
\begin{equation}
    \Delta_C \Delta \omega_C \geq \frac{1}{2}.
\end{equation}
The market dependence we want to resolve has characteristic rates.  The two basic ones are the shared-refresh or clock-overlap rate $\lambda^C_{ij}$ and the coupling-response rate $\kappa^C_{ij}$, both in units of inverse clock $C$ \citep{ChangPienaarGebbie2021PhysicaA,AngstmannGebbie2026LOB}.  Set
\begin{equation}
    \beta^C_{ij}
    :=
    \min\{\lambda^C_{ij},\kappa^C_{ij}\}.
\end{equation}
The windowed signal cannot resolve that dependence scale unless its spectral uncertainty is smaller than the slow rate being measured:
\begin{equation}
    \Delta \omega_C \lesssim \beta^C_{ij}.
\end{equation}
Combining this requirement with the Gabor inequality gives a minimum window:
\begin{equation}
    \Delta_C
    \gtrsim
    \frac{1}{2\beta^C_{ij}}
    =
    \frac{1}{2\min\{\lambda^C_{ij},\kappa^C_{ij}\}}.
\end{equation}
This is the resolution floor.  Below it, a short-window correlation estimate is not only noisy.  It is under-resolved.  The window is too narrow to distinguish the rate at which cross-asset dependence emerges.

The Gabor bound should therefore be understood as a minimum resolution floor: below this scale the relevant market-rate structure is not yet identifiable from the windowed signal.  The order-book calculation below makes this less philosophical.  It shows how a weighted imbalance functional inherits a response rate from the signed order-density dynamics.  The reaction--diffusion model supplies the rate.  The Gabor inequality supplies the minimum window over which that rate can be resolved.  For practical trading, however, one usually needs a more conservative maturity threshold.  That threshold is constructed later from the Epps attenuation functions, which describe how realised correlation approaches its large-window value after the Gabor floor has been exceeded.

For imbalance-based toy signals we include one more rate.  Let \(\omega_I^C\) be an indicative imbalance-response or reversal frequency in clock \(C\).  The slow rate used in the toy Gabor floor is then
\begin{equation}
    \beta_I^C
    =
    \min\{\lambda^C_{ij},\kappa^C_{ij},\omega_I^C\}.
\end{equation}
The following examples are calendar-time examples, so \(C=t\) and the rates are measured per second.  The numbers are not calibrated empirical estimates.  They are stylised dimensional checks showing how the principle would be used once desk-specific estimates of \(\lambda^C_{ij}\), \(\kappa^C_{ij}\), and \(\omega_I^C\) are available.

\begin{table}[H]
\centering
\small
\resizebox{\linewidth}{!}{%
\begin{tabular}{lccccc}
\hline
Pair type
&
\(\lambda^C_{ij}\)
&
\(\kappa^C_{ij}\)
&
\(\omega_I^C\)
&
\(\beta_I^C\)
&
\(\Delta_C^{\rm Gabor}\)
\\
\hline
Liquid pair
&
\(12\,{\rm s}^{-1}\)
&
\(2\,{\rm s}^{-1}\)
&
\(5\,{\rm s}^{-1}\)
&
\(2\,{\rm s}^{-1}\)
&
\(0.25\,{\rm s}\)
\\
Less liquid pair
&
\(0.5\,{\rm s}^{-1}\)
&
\(0.15\,{\rm s}^{-1}\)
&
\(0.3\,{\rm s}^{-1}\)
&
\(0.15\,{\rm s}^{-1}\)
&
\(3.3\,{\rm s}\)
\\
\hline
\end{tabular}%
}
\caption{Stylised Gabor resolution floors for weighted imbalance in calendar time.  The entries are not calibrated estimates; they are dimensional examples.  The shared-refresh rate \(\lambda^C_{ij}\), cross-asset response rate \(\kappa^C_{ij}\), and imbalance-response or reversal frequency \(\omega_I^C\) are all measured in inverse seconds.  The slowest rate \(\beta_I^C=\min\{\lambda^C_{ij},\kappa^C_{ij},\omega_I^C\}\) binds because the signal cannot mature faster than its slowest required mechanism.  Thus \(\Delta_C^{\rm Gabor}\gtrsim 1/(2\beta_I^C)\).  In the liquid example, \(\beta_I^C=2\,{\rm s}^{-1}\), so the floor is \(1/(2\times2)=0.25\) seconds: a \(50\)--\(100\) millisecond statistic may see the book, but it is too short to resolve the slow cross-asset response.  In the less liquid example, \(\beta_I^C=0.15\,{\rm s}^{-1}\), so the floor is about \(3.3\) seconds.  The time--frequency resolution floor follows \citet{Gabor1946}; the market rates are motivated by the clock and coupling mechanisms in \citet{ChangPienaarGebbie2021PhysicaA} and \citet{AngstmannGebbie2026LOB}.}
\label{tab:gabor-toy-scales}
\end{table}

\paragraph{Example: Gabor floor}
The point is not that the Gabor floor is a trading horizon.  It is only the first scale at which the slowest response rate can begin to be resolved.  Sub-second imbalance estimates may still be useful for local execution, but as cross-asset signals they can be under-resolved -- this is important.

\section{Order flow as the signal}

One should start with the order flow, but with a distinction between the clock and the measured field. Unsigned activity defines a clock. Signed activity defines an observable on that clock.  For trades at event times $t_n$, with signs $\varepsilon_n\in\{-1,+1\}$ and sizes $v_n$ \citep{angstmann2026LMF}, define signed order flow by
\begin{equation}
    Q(t)=\sum_{t_n\leq t}\varepsilon_n v_n.
\end{equation}
In calendar time a windowed imbalance may be written as
\begin{equation}
    I_{\Delta_t}(t)=
    \frac{\sum_{t_n\in(t,t+\Delta_t]}\varepsilon_n v_n}
    {\sum_{t_n\in(t,t+\Delta_t]}v_n}.
\end{equation}
More generally, if $\tau$ denotes the coordinate of the chosen clock $C$, the imbalance over the window is best written as $I^C_{\Delta_C}(\tau)$. By contrast, the unsigned event-count and volume clocks are
\begin{equation}
    N(t)=\sum_{t_n\leq t}1, \qquad
    V(t)=\sum_{t_n\leq t}v_n.
\end{equation}
The Fourier dual of a chosen clock $C$ is then a frequency per unit of that clock.  Formally, after the market data have been reparametrised by that clock,
\begin{equation}
    \Delta_t\,\Delta\omega_t \geq \frac{1}{2}, \qquad
    \Delta_N\,\Delta\omega_N \geq \frac{1}{2}, \qquad
    \Delta_V\,\Delta\omega_V \geq \frac{1}{2}.
\end{equation}
Here $\omega_N$ is a frequency per event, while $\omega_V$ is a frequency per unit volume.  This is only a clock-coordinate statement.  The financial content comes from the signal one puts on that clock: signed order flow $Q$, imbalance $I$, depth imbalance, spread, or a pair-spread process.  A natural energy-like quantity for imbalance is its liquidity-weighted spectral power.  For a chosen clock \(C\), define
\begin{equation}
    \mathcal{E}_{I}^{C}
    =
    \int
    \Lambda^{C}(\omega_C)
    \left|\widehat{I}^{C}(\omega_C)\right|^2
    \,d\omega_C ,
    \label{eq:imbalance-energy}
\end{equation}
where \(\Lambda^{C}(\omega_C)\) is a liquidity, impact, or execution-cost weighting kernel.  This quantity is energy-like in the signal-processing sense: it measures how much imbalance power is present at each frequency of the chosen market clock, with frequencies that are more costly or more price-relevant receiving greater weight.

The Gabor limit applies to the windowed imbalance signal, and the energy estimate inherits the resulting resolution limit.  In practice the imbalance signal is never observed over an infinite interval.  It is observed through a finite window $\Delta_C$.  The spectrum is therefore blurred.  If $I^C_{\Delta_C}$ denotes the imbalance observed through such a window, then
\begin{equation}
    \Delta_C \Delta\omega_I^C \geq \frac{1}{2}.
\end{equation}
The finite-window estimate
\begin{equation}
    \widehat{\mathcal{E}}_{I}^{C}(\Delta_C)
    =
    \int
    \Lambda^{C}(\omega_C)
    \left|
        \widehat{I}^{C}_{\Delta_C}(\omega_C)
    \right|^2
    \,d\omega_C
\end{equation}
cannot localise the energy of the windowed imbalance signal arbitrarily sharply in both clock time and frequency.  Short windows give good event localisation.  They give poor rate localisation.  This is exactly the Gabor trade-off, inherited by the imbalance-energy estimate \citep{Gabor1946}.

This becomes operational once the relevant imbalance response rate is identified.  Suppose the weighted imbalance is intended to resolve a mechanism with characteristic rate
\begin{equation}
    \beta_{I}^{C}
    =
    \min\{
        \lambda_{ij}^{C},
        \kappa_{ij}^{C},
        \omega_{I}^{C}
    \},
\end{equation}
where $\lambda_{ij}^{C}$ is the shared-refresh or clock-overlap rate, $\kappa_{ij}^{C}$ is the cross-asset coupling-response rate, and $\omega_I^C$ is the characteristic imbalance-reversal frequency in clock $C$.  Resolving this mechanism requires
\begin{equation}
    \Delta\omega_I^C \lesssim \beta_I^C .
\end{equation}
The Gabor inequality then gives
\begin{equation}
    \Delta_C
    \gtrsim
    \frac{1}{2\beta_I^C}.
\end{equation}
Below this window, $\widehat{\mathcal{E}}_{I}^{C}(\Delta_C)$ is under-resolved.  The estimator may see local imbalance.  It cannot yet tell whether that imbalance belongs to a persistent cross-asset response, a fast liquidity-replenishing oscillation, or a transient event-level fluctuation.

For a book-level implementation, one may take the imbalance itself to be a weighted functional of the signed order density near the transaction boundary.  This follows the reaction--diffusion order-book picture, where the transaction price is the moving zero of the signed density \citep{AngstmannGebbie2026LOB}:
\begin{equation}
    I_{w}^{(j),C}(\tau)
    =
    \frac{
    \int_{\mathbb R}
    w(y)\phi_{C}^{(j)}(p_j^C(\tau)+y,\tau)\,dy
    }{
    \int_{\mathbb R}
    w(y)\left|\phi_{C}^{(j)}(p_j^C(\tau)+y,\tau)\right|\,dy
    } .
\end{equation}
Here $y=x-p_j^C(\tau)$, where $p_j^C(\tau)$ is the transaction boundary viewed in clock $C$.  The weight $w$ concentrates the estimate near the best quotes or the first few price levels.  The clock label $C$ means that the field has been viewed in the chosen clock.  If the denominator is close to zero, the imbalance is not well-defined.  In applications one should either discard that state or add a small regularising depth.

The order-book dynamics give the response scale.  In the coupled reaction--diffusion approximation,
\begin{equation}
\partial_t\phi^{(j)}
=
D_\alpha\partial_x^2(D_t^{1-\alpha}\phi^{(j)})
-
\nu_j\phi^{(j)}
+
c^{(j,k)}(x,t),
\end{equation}
Here $D_\alpha$ is the operational-time diffusion scale of the order-density field, with $\alpha=1$ giving ordinary diffusion and $0<\alpha<1$ giving the fractional or subdiffusive case.  The parameter $\nu_j$ is the cancellation or annihilation rate in book $j$, so it sets one local relaxation scale for the signed order density.  The term $c^{(j,k)}$ collects order creation, shocks, and cross-book coupling terms \citep{AngstmannGebbie2026LOB}.  This equation is written in the model's operational time; the clock label $C$ in the imbalance functional indicates the observation parametrisation used by the estimator.  If $w$ has spatial width $\ell_w$, then the imbalance mainly probes price-space modes $k_w\sim \ell_w^{-1}$.  For ordinary diffusion, $\alpha=1$, the corresponding response-rate scale is
\begin{equation}
    \beta_{I,w}^{(j),C}
    \approx
    \nu_j + D_1 k_w^2 .
\end{equation}
For $0<\alpha<1$, the relaxation is not a single exponential.  A useful dimensional proxy is
\begin{equation}
    \beta_{I,w}^{(j),C}
    \approx
    \nu_j + (D_\alpha k_w^2)^{1/\alpha} .
\end{equation}
This should be read only as a dimensional effective rate, not as the exact relaxation spectrum or a single pole of the fractional dynamics.  The fractional relaxation is broad and history-dependent.

For a purely local weighted-imbalance diagnostic, this gives the corresponding local Gabor floor.  For a cross-asset imbalance signal, however, the relevant effective rate must also include shared refresh and cross-book coupling:
\begin{equation}
    \beta_{\mathrm{eff}}^C
    =
    \min\{
        \lambda_{ij}^C,
        \kappa_{ij}^C,
        \beta_{I,w}^{(j),C}
    \}.
\end{equation}
The Gabor resolution floor for estimating weighted imbalance at this spatial scale is therefore
\begin{equation}
    \Delta_C
    \gtrsim
    \frac{1}{2\beta_{\mathrm{eff}}^C},
\end{equation}
with the local-book special case obtained by replacing $\beta_{\mathrm{eff}}^C$ with $\beta_{I,w}^{(j),C}$.
A narrower book-weight $w$ gives a more local imbalance estimate.  It also pushes the estimate toward higher spatial and temporal response rates.  The uncertainty principle makes the trade-off explicit: local imbalance can be measured quickly, but stable imbalance energy $\mathcal{E}_{I}^{C}$ at the relevant market-response scale requires a longer observation window. A trader can measure the local book imbalance almost immediately.  What cannot be known immediately is whether that imbalance is just a queue-level flicker, or whether it is persistent pressure that will move prices or transmit signal to another asset.  That second object needs a longer window. The market needs time, trades, and shared refreshes before the imbalance becomes a reliable cross-asset signal -- this is because markets are event-driven.

\section{Epps attenuation as a resolution function}

The Epps effect is probably the empirical and modelling situation where this uncertainty is most directly visible \citep{Epps1979,TothKertesz2009}.  Building on stochastic business-time ideas \citep{Clark1973}, \citet{ChangPienaarGebbie2021PhysicaA} studied alternative sampling schemes to show that the Epps curve depends on the chosen clock: calendar time, trade time, and volume time need not produce the same emergence profile for correlation. This motivates treating realised correlation as a scale- and clock-dependent observable rather than as a noisy estimate of a unique instantaneous object. This perspective complements the established non-synchronous covariance and Fourier-estimation literatures \citep{HayashiYoshida2005,MalliavinMancino2002}.

For a chosen clock $C$, define returns over a clock-window $\Delta_C$ by
\begin{equation}
    R^{(j),C}_{\Delta_C}(\tau)
    = p_j^C(\tau+\Delta_C)-p_j^C(\tau),
\end{equation}
and the corresponding realised correlation by
\begin{equation}
    \rho^C_{\Delta_C}
    =
    \operatorname{Corr}\!
    \left(
        R^{(1),C}_{\Delta_C},
        R^{(2),C}_{\Delta_C}
    \right).
\end{equation}
In general,
\begin{equation}
    \rho^t_{\Delta_t}
    \neq
    \rho^N_{\Delta_N}
    \neq
    \rho^V_{\Delta_V},
\end{equation}
because changing the clock changes the aggregation and synchronisation convention.

A simple analytic form arises from asynchronous refresh clocks \citep{TothKertesz2009,AngstmannGebbie2026LOB}:
\begin{equation}
    \rho^{C,\mathrm{sub}}_{\Delta_C}
    \simeq
    \rho^{C}_{\infty}
    \left[
        1-\frac{1-e^{-\lambda^{C}_{ij}\Delta_C}}{\lambda^{C}_{ij}\Delta_C}
    \right].
\end{equation}
A similar form arises from finite coupling response in coupled order books:
\begin{equation}
    \rho^{C,\mathrm{coup}}_{\Delta_C}
    \simeq
    \rho^{C}_{\infty}
    \left[
        1-\frac{1-e^{-\kappa^{C}_{ij}\Delta_C}}{\kappa^{C}_{ij}\Delta_C}
    \right],
\end{equation}
where $\kappa^{C}_{ij}$ is the relaxation rate by which coupling-induced order flow moves the relevant price or reaction boundary in clock $C$ \citep{AngstmannGebbie2026LOB}.  A first-order combined approximation is
\begin{equation}
    \rho^{C,\mathrm{comb}}_{\Delta_C}
    \simeq
    \rho^{C}_{\infty}
    \left[
        1-\frac{1-e^{-\lambda^{C}_{ij}\Delta_C}}{\lambda^{C}_{ij}\Delta_C}
    \right]
    \left[
        1-\frac{1-e^{-\kappa^{C}_{ij}\Delta_C}}{\kappa^{C}_{ij}\Delta_C}
    \right].
\end{equation}
This factorisation is a leading-order modelling approximation.  It is most natural when clock overlap and local coupling response can be treated as approximately separable \citep{AngstmannGebbie2026LOB}.

The function
\begin{equation}
    f(x)=1-\frac{1-e^{-x}}{x} \label{eq:attenuation}
\end{equation}
is a resolution function.  It satisfies
\begin{equation}
    f(x)\sim \frac{x}{2}, \qquad x\downarrow 0,
\end{equation}
and tends to one as $x\to\infty$.  Hence short-window correlation is suppressed not only because of estimator noise, but because the measurement window may not contain enough shared activity or enough response time for dependence to emerge.

This gives the proposed trading principle:
\begin{quote}
\emph{At event scales, the more precisely one localises market activity in time, the less well one can resolve stable cross-asset dependence. The more one resolves dependence, the more one has coarse-grained away the event-time structure that generated it.}
\end{quote}

\section{Six practical trading rules}

The following rules follow from the principle.

\begin{enumerate}[label=\textbf{Rule \arabic*.}, leftmargin=*]
    \item \textbf{Estimate a correlation maturity horizon.}  For each pair $i,j$ and clock $C$, define
    \begin{equation}
        \Delta_{ij,C}^{q}
        =
        \inf\{\Delta_C:\rho^C_{ij}(\Delta_C)\geq q\rho_{ij}^{C,\infty}\},
    \end{equation}
    for a threshold such as $q=0.8$ or $q=0.9$.  Below, \(\Delta_C^{80}\) and \(\Delta_C^{90}\) denote the corresponding approximate maturity horizons for \(q=0.8\) and \(q=0.9\).  A correlation signal should not be treated as operationally mature below this horizon.

    \item \textbf{Do not treat low short-horizon correlation as diversification.}  A low value of $\rho^C_{\Delta_C}$ may indicate independence, but it may also indicate that dependence has not yet emerged, or has vanished, under the chosen clock.  Short-horizon diversification should therefore be stress-tested against the plateau value $\rho^{C}_{\infty}$.

    \item \textbf{Execute in market-activity time, but account in calendar time.}  Execution problems are naturally controlled by event time, volume time, queue time, or dollar-volume time.  Portfolio reporting, funding, and client risk are usually calendar-time objects.  Clock mismatch is therefore a real implementation risk, particularly when accounting for borrowing costs or currency conversion.

    \item \textbf{Monitor refresh and response rates.}  If $\lambda^C_{ij}$ falls, common clock overlap deteriorates.  If trade durations shorten or lengthen materially, price impact and liquidity conditions may also change \citep{DufourEngle2000}.  If $\kappa^C_{ij}$ falls, coupling is slower and hedges mature later.  When either rate deteriorates, traders should probably reduce leverage, widen holding horizons, or require stronger signals.

    \item \textbf{Charge for clock mismatch.}  A hedge built on one clock and executed on another carries clock-basis risk. Almost all trades are on at least two clocks. This risk is especially relevant near opens, closes, auctions, index rebalances, news events, liquidity holes, and stress episodes.

    \item \textbf{Use spectral order-flow diagnostics.}  The spectrum of signed order flow or imbalance can reveal whether the market is dominated by slow persistent pressure or fast oscillatory liquidity provision.  The same calendar interval can contain very different market information depending on the order-flow spectrum.
\end{enumerate}

A useful numerical guide follows heuristically from the resolution function in Equation \ref{eq:attenuation}: $f(5)\approx 0.80$ and $f(10)\approx 0.90$.  If a single attenuation mechanism dominates, one needs approximately
\begin{equation}
    \Delta_C \gtrsim \frac{5}{\lambda^{C}_{ij}}
    \quad\text{or}\quad
    \Delta_C \gtrsim \frac{5}{\kappa^{C}_{ij}}
\end{equation}
to resolve about eighty percent of the long-run dependence in whichever clock one is currently using: calendar seconds $t$, number of trades $N$, or volume time $V$.  If clock overlap and coupling response both matter, a simple two-mechanism screen is
\begin{equation}
    \Delta_C \gtrsim
    \max\left\{\frac{10}{\lambda^{C}_{ij}},\frac{10}{\kappa^{C}_{ij}}\right\}.
\end{equation}
This puts each separate attenuation factor near ninety percent.  The product is then still only about eighty percent, so a desk that wants a higher combined threshold should choose a larger multiplier.  The units of this rule are the units of $C$: seconds or minutes for calendar time, event counts for event time, and shares, contracts, or value traded for volume time.

For pure correlation maturity, the two rates in the screen are \(\lambda^C_{ij}\) and \(\kappa^C_{ij}\).  For the imbalance examples below we use the three-rate version, because the signal must also wait for the local imbalance response or reversal rate \(\omega_I^C\).  Thus
\begin{align}
    \Delta_C^{80}
    &\approx
    \max
    \left\{
        \frac{5}{\lambda^C_{ij}},
        \frac{5}{\kappa^C_{ij}},
        \frac{5}{\omega_I^C}
    \right\},
    \\
    \Delta_C^{90}
    &\approx
    \max
    \left\{
        \frac{10}{\lambda^C_{ij}},
        \frac{10}{\kappa^C_{ij}},
        \frac{10}{\omega_I^C}
    \right\}.
\end{align}
The max is deliberately blunt.  The trader waits for shared refresh, cross-book response, and local imbalance reversal.  The slowest one binds.

\begin{table}[H]
\centering
\small
\resizebox{\linewidth}{!}{%
\begin{tabular}{lcccc}
\hline
Pair type
&
\(\Delta_C^{\rm Gabor}\)
&
\(\Delta_C^{80}\)
&
\(\Delta_C^{90}\)
&
Binding rate
\\
\hline
Liquid pair
&
\(0.25\,{\rm s}\)
&
\(2.5\,{\rm s}\)
&
\(5.0\,{\rm s}\)
&
\(\kappa^C_{ij}=2\,{\rm s}^{-1}\)
\\
Less liquid pair
&
\(3.3\,{\rm s}\)
&
\(33.3\,{\rm s}\)
&
\(66.7\,{\rm s}\)
&
\(\kappa^C_{ij}=0.15\,{\rm s}^{-1}\)
\\
\hline
\end{tabular}%
}
\caption{Stylised Epps maturity scales using the same rates as Table~\ref{tab:gabor-toy-scales}.  The Gabor floor is the first resolvability scale; it is not yet a mature trading horizon.  The Epps thresholds are deliberately more conservative because the attenuation curve \(f(x)=1-(1-e^{-x})/x\) approaches its plateau only after several refresh or response times.  For a single mechanism, \(f(5)\approx0.80\) and \(f(10)\approx0.90\), so the corresponding windows are about \(5/\beta\) and \(10/\beta\).  In the liquid example the binding rate is \(2\,{\rm s}^{-1}\), giving \(2.5\) and \(5.0\) seconds.  In the less liquid example the binding rate is \(0.15\,{\rm s}^{-1}\), giving about \(33.3\) and \(66.7\) seconds.  The intuition is that a signal can become barely resolvable before it is mature enough to trade as stable cross-asset dependence.  The attenuation interpretation follows the Epps effect literature and its coupled order-book decomposition \citep{Epps1979,ChangPienaarGebbie2021PhysicaA,AngstmannGebbie2026LOB}.}
\label{tab:epps-toy-scales}
\end{table}

\paragraph{Example: Epps attenuation}
The useful trading scale is much longer than the Gabor floor.  The market needs several shared-refresh or response times before a pair signal should be treated as mature.  Otherwise a one-second pair signal may mean very little -- the algorithm is chasing ghosts.

\section{Implication for fund managers}

For fund managers this is really a model-governance lesson, rather than being about high-frequency prediction.  Covariance matrices should be estimated at the decision horizon.  Intraday execution risk, daily portfolio risk, monthly allocation, and quarterly strategic allocation do not necessarily use the same covariance object.  A factor or hedge that is stable in calendar time but unstable in volume time may be a liquidity-seasonality effect.  Conversely, a relation that is visible in event time but weak in calendar time may be an order-flow effect. So what one is trying to do really matters when choosing the clock.

The broader interpretation is consistent with the idea that market time is not unique \citep{AngstmannGebbie2026NonUnique}.  If event-to-continuum maps are not canonical, then hedging, covariance measurement, and pricing are attached to a chosen representation rather than to a unique underlying clock.  Coarse-grained calendar-time models can still be useful, but their usefulness should be understood as emergent and horizon-dependent.

\section{Discussion and conclusion}
The proposed Gabor--Epps uncertainty principle is best understood as a practical resolution principle. Correlation is not observed in time alone.  It is resolved through market activity, clock overlap, and order-flow response.  Before trading correlation, estimate its Epps curve; before hedging, estimate its maturity horizon; before allocating risk, check whether the dependence is stable across clocks. The basic idea here is to use Fourier/Gabor reasoning, stochastic clocks, and Epps-type correlation emergence to make transparent a practical fact: high-frequency dependence is a property of the measurement clock and the order-flow mechanism as much as it is a property of the assets themselves.

A final distinction is important.  The Gabor floor is only a minimum resolvability scale: below \(1/(2\beta)\), the relevant rate is not cleanly identifiable from the finite-window signal.  It is not a recommendation to trade at that scale.  The Epps maturity horizon is more conservative: it asks how many refresh or response times are needed before realised dependence is close enough to its plateau to be operationally useful.  In the liquid toy example, the Gabor floor is \(0.25\) seconds, while an eighty percent maturity screen is closer to \(2.5\) seconds.  In the less liquid example, these become roughly \(3.3\) seconds and \(33.3\) seconds.  That gap is the point: first resolvability and trading maturity are not the same object.

The safe claim is thus: in any finite-window market measurement, there is a time--rate resolution trade-off.  When the rate being estimated is a market-dependence rate -- shared refresh, coupling response, or imbalance reversal -- this trade-off imposes a minimum useful observation scale, and the Epps effect then gives a more conservative maturity scale for correlation trading.  This is not a theorem of empirical fact and not a fundamental law of markets.  It should also not be confused with popular or quantum-mechanical notions of uncertainty; that would be pseudoscience in this context.  The intended claim is a signal-processing and market-microstructure claim about finite windows, market clocks, and rate-limited dependence.  The intention is to provide a useful signal-processing intuition for finite-window dependence estimates while connecting that intuition to order-book dynamics and the Epps effect.

\section*{Acknowledgements}

I thank Lionel Yelibi, Alexa Orton, Daniella Stevenson and Chris Angstmann for various conversations about trading analytics, strategies and time -- all remaining errors are mine.

\section*{Declaration of generative AI}

During the preparation of this work I used ChatGPT 5.5 for proof reading, reviewing and checking. I reviewed and edited the output as needed and take full responsibility for the content.

\section*{Declaration of competing interest}
The author declares that he has no known financial or non-financial competing interests that could have appeared to influence the work reported in this manuscript.  The author has no employment, consultancy, shareholding, honorarium, patent, paid expert testimony, or other financial relationship connected to the claims or examples in this paper.  The examples are stylised and are included for scientific and pedagogical illustration only.

\section*{Disclaimer}
This paper is for research and educational purposes only.  It is not investment advice, trading advice, risk-management advice, or a recommendation to buy, sell, short, hedge, or hold any financial instrument.  The rules of thumb and numerical examples are stylised and are not calibrated trading signals.  They should not be used for live trading or portfolio allocation without independent estimation, validation, stress testing, and appropriate risk controls.

Any use of the ideas, formulae, terminology, or rules of thumb in this paper is at the user's own risk.  The author accepts no responsibility for financial losses, trading losses, model failures, operational losses, or other damages arising directly or indirectly from such use.

\bibliography{GaborEppsUncertainty-v2.3}

\end{document}